\documentclass[a4paper,11pt,preprint]{elsarticle}

\pdfoutput=1 
\usepackage{amssymb}
\usepackage{amsmath}

\usepackage[T1]{fontenc} 
\usepackage[lmargin=2.65cm, rmargin=2.65cm, bmargin=3.5cm]{geometry}

\begin{document} 

\begin{frontmatter}
\title{One-particle reducible contribution to the one-loop spinor propagator  in a constant field}	

\author[a]{N. Ahmadiniaz}
\ead{ahmadiniaz@ibs.re.kr}

\author[b,c]{F. Bastianelli}
\ead{fiorenzo.bastianelli@bo.infn.it} 

\author[d,c]{O. Corradini}
\ead{olindo.corradini@unimore.it}

\author[e]{J.P. Edwards}
\ead{jedwards@ifm.umich.mx}

\author[e]{C. Schubert\corref{cor1}}
\ead{schubert@ifm.umich.mx}

\address[a]{Center for Relativistic Laser Science, Institute for Basic Science, 61005 Gwangju, Korea}
\address[b]{Dipartimento di Fisica ed Astronomia, Universit\`a di Bologna, Via Irnerio 46, I-40126 Bologna, Italy}
\address[c]{INFN, Sezione di Bologna, Via Irnerio 46, I-40126 Bologna, Italy}
\address[d]{Dipartimento di Scienze Fisiche, Informatiche e Matematiche,
 Universit\`a degli Studi di Modena e Reggio Emilia, Via Campi 213/A, I-41125 Modena, Italy}
\address[e]{Instituto de F\'isica y Matem\'aticas
Universidad Michoacana de San Nicol\'as de Hidalgo
Edificio C-3, Apdo. Postal 2-82
C.P. 58040, Morelia, Michoac\'an, Mexico}

\cortext[cor1]{Corresponding author.}


\begin{abstract}
Extending work by Gies and Karbstein on the Euler-Heisenberg Lagrangian, it has recently been shown that the one-loop propagator of  
a charged scalar particle in a constant electromagnetic field has a one-particle reducible contribution in addition to the well-studied irreducible one.
Here we further generalize this result to the spinor case, and find the same relation between the reducible term, the tree-level propagator and the
one-loop Euler-Heisenberg Lagrangian as in the scalar case. Our demonstration uses a novel worldline path integral 
representation of the photon-dressed spinor propagator in a constant electromagnetic field background.
\end{abstract}

\end{frontmatter}

\tableofcontents

\def\green{\color{green}}
\def\blue{\color{blue}}
\def\red{\color{red}}
\def\black{\color{black}}
%

\def\veps{\varepsilon}
\newcommand{\Zz}{\mathcal{Z}}
\newcommand{\Zzp}{\mathcal{Z}^{\prime}}
\newcommand{\detZ}{\textrm{det}^{-\frac{1}{2}}\left[\frac{\sin\Zz}{\mathcal{\Zz}}\right]}
\newcommand{\detZp}{\textrm{det}^{-\frac{1}{2}}\left[\frac{\sin \Zzp}{\Zzp}\right]}
\newcommand{\detZs}{\textrm{det}^{-\frac{1}{2}}\left[\frac{\tan\Zz}{\mathcal{\Zz}}\right]}
\newcommand{\detZps}{\textrm{det}^{-\frac{1}{2}}\left[\frac{\tan\Zzp}{\Zzp}\right]}
\newcommand{\pdetZ}{\textrm{det}^{-\frac{1}{2}}\left[\cos \Zz \right]}
\newcommand{\pdetZp}{\textrm{det}^{-\frac{1}{2}}\left[\cos(\Zzp)\right]}

\newcommand{\tZz}{\frac{\tan\Zz}{\Zz}}
\newcommand{\tZzp}{\frac{\tan\Zzp}{\Zzp}}
\newcommand{\link}{\Big\vert_k}

\newcommand{\xm}{x_{-}}
\newcommand{\xp}{x_{+}}
\newcommand{\yp}{y_{+}}
\newcommand{\ym}{y_{-}}

\newcommand{\delC}{\underset{\smile}{\Delta}}
\newcommand{\ddelC}{{^{\bullet}\!\delC}}
\newcommand{\delCd}{{\delC\!^{\bullet}}}
\newcommand{\ddelCd}{{^{\bullet}\!\delC\!^{\bullet}}}
\newcommand{\odelC}{{^{\circ}\!\delC}}
\newcommand{\delCo}{{\delC\!^{\circ}}}
\newcommand{\odelCo}{{^{\circ}\!\delC\!^{\circ}}}
\newcommand{\odelCd}{{^{\circ}\!\delC\!^{\bullet}}}
\newcommand{\ddelCo}{{^{\bullet}\!\delC\!^{\circ}}}

\newcommand{\gb}{{\mathcal{G}_{B}}}
\newcommand{\gbd}{{\dot{\mathcal{G}}_{B}}}
\newcommand{\gbdm}{\dot{\mathcal{G}}_{B \mu\nu}}
\newcommand{\gf}{{\mathcal{G}_{F}}}
\newcommand{\gfd}{{\dot{\mathcal{G}}_{F}}}
\newcommand{\gfdm}{\dot{\mathcal{G}}_{F \mu\nu}}

\def\cZ{{\cal Z}}

\def\cosech{\rm cosech}
\def\sech{\rm sech}
\def\coth{\rm coth}
\def\tanh{\rm tanh}
\def\tan{\rm tan}
\def\half{{1\over 2}}
\def\third{{1\over3}}
\def\fourth{{1\over4}}
\def\fifth{{1\over5}}
\def\sixth{{1\over6}}
\def\seventh{{1\over7}}
\def\eigth{{1\over8}}
\def\ninth{{1\over9}}
\def\tenth{{1\over10}}
\def\conj{{{\rm c.c.}}}
\def\bN{\mathop{\bf N}}
\def\R{{\rm I\!R}}
\def\Eins{{\mathchoice {\rm 1\mskip-4mu l} {\rm 1\mskip-4mu l}
{\rm 1\mskip-4.5mu l} {\rm 1\mskip-5mu l}}}
\def\Z{{\mathchoice {\hbox{$\sf\textstyle Z\kern-0.4em Z$}}
{\hbox{$\sf\textstyle Z\kern-0.4em Z$}}
{\hbox{$\sf\scriptstyle Z\kern-0.3em Z$}}
{\hbox{$\sf\scriptscriptstyle Z\kern-0.2em Z$}}}}
\def\abs#1{\left| #1\right|}
\def\com#1#2{
        \left[#1, #2\right]}
\def\contract{\makebox[1.2em][c]{
        \mbox{\rule{.6em}{.01truein}\rule{.01truein}{.6em}}}}
\def\ltap{\ \raisebox{-.4ex}{\rlap{$\sim$}} \raisebox{.4ex}{$<$}\ }
\def\gtap{\ \raisebox{-.4ex}{\rlap{$\sim$}} \raisebox{.4ex}{$>$}\ }
\def\mn{{\mu\nu}}
\def\rs{{\rho\sigma}}
\newcommand{\Det}{{\rm Det}}
\def\Tr{{\rm Tr}\,}
\def\tr{{\rm tr}\,}
\def\sumij{\sum_{i<j}}
\def\e{\,{\rm e}}
\def\br{{\bf r}}
\def\bp{{\bf p}}
\def\bq{{\bf q}}
\def\bx{{\bf x}}
\def\by{{\bf y}}
\def\brhat{{\bf \hat r}}
\def\bv{{\bf v}}
\def\ba{{\bf a}}
\def\bE{{\bf E}}
\def\bB{{\bf B}}
\def\bA{{\bf A}}
\def\b0{{\bf 0}}
\def\pa{\partial}
\def\dA{\partial^2}
\def\ddx{{d\over dx}}
\def\ddt{{d\over dt}}
\def\der#1#2{{d #1\over d#2}}
\def\lie{\hbox{\it \$}} 
\def\partder#1#2{\frac{\partial #1}{\partial #2}}
\def\secder#1#2#3{{\partial^2 #1\over\partial #2 \partial #3}}
%
\def\be{\begin{equation}}
\def\ee{\end{equation}\noindent}
\def\bear{\begin{eqnarray}}
\def\ear{\end{eqnarray}\noindent}
\def\bec{\blue\begin{equation}}
\def\eec{\end{equation}\black\noindent}
\def\bearc{\blue\begin{eqnarray}}
\def\earc{\end{eqnarray}\black\noindent}
\def\benn{\begin{enumerate}}
\def\enn{\end{enumerate}}
\def\veject{\vfill\eject}
\def\ven{\vfill\eject\noindent}
%
\def\eq#1{{eq. (\ref{#1})}}
\def\eqs#1#2{{eqs. (\ref{#1}) -- (\ref{#2})}}
%
\def\inv#1{\frac{1}{#1}}
\def\sumninf{\sum_{n=0}^{\infty}}
%
\def\totint{\int_{-\infty}^{\infty}}
\def\posint{\int_0^{\infty}}
\def\negint{\int_{-\infty}^0}
\def\pint{{\dps\int}{dp_i\over {(2\pi)}^d}}
\def\intdp3{\int\frac{d^3p}{(2\pi)^3}}
\def\intdp4{\int\frac{d^4p}{(2\pi)^4}}
\def\scalprop#1{\frac{-i}{#1^2+m^2-i\epsilon}}
%
\newcommand{\GeV}{\mbox{GeV}}
\def\FFdual{F\cdot\tilde F}
\def\bra#1{\langle #1 |}
\def\ket#1{| #1 \rangle}
\def\braket#1#2{\langle {#1} \mid {#2} \rangle}
\def\vev#1{\langle #1 \rangle}
\def\matel#1#2#3{\langle #1\mid #2\mid #3 \rangle}
\def\rightvac{\mid0\rangle}
\def\leftvac{\langle0\mid}
\def\ihbar{{i\over\hbar}}
\def\lagr{{\cal L}}
\def\sigmabar{{\bar\sigma}}
\def\ge{\hbox{$\gamma_1$}}
\def\gz{\hbox{$\gamma_2$}}
\def\gd{\hbox{$\gamma_3$}}
\def\go{\hbox{$\gamma_1$}}
\def\gt{\hbox{$\gamma_2$}}
\def\gth{\hbox{$\gamma_3$}} 
\def\gf{\hbox{$\gamma_5\;$}}
\def\slash#1{#1\!\!\!\raise.15ex\hbox {/}}
\newcommand{\slD}{\,\raise.15ex\hbox{$/$}\kern-.27em\hbox{$\!\!\!D$}}
\newcommand{\slpartial}{\raise.15ex\hbox{$/$}\kern-.57em\hbox{$\partial$}}
\newcommand{\PP}{\cal P}
\newcommand{\G}{{\cal G}}
\newcommand{\nc}{\newcommand}
\nc{\spa}[3]{\left\langle#1\,#3\right\rangle}
\nc{\spb}[3]{\left[#1\,#3\right]}
\nc{\ksl}{\not{\hbox{\kern-2.3pt $k$}}}
\nc{\hf}{\textstyle{1\over2}}
\nc{\pol}{\varepsilon}
\nc{\tq}{{\tilde q}}
\nc{\esl}{\not{\hbox{\kern-2.3pt $\pol$}}}
\newcommand{\cL}{\cal L}
\newcommand{\D}{\cal D}
\newcommand{\Dhalf}{{D\over 2}}
\def\eps{\epsilon}
\def\epshalf{{\epsilon\over 2}}
\def\lag{( -\partial^2 + V)}
\def\freeexp{{\rm e}^{-\int_0^Td\tau {1\over 4}\dot x^2}}
\def\kinb{{1\over 4}\dot x^2}
\def\kinf{{1\over 2}\psi\dot\psi}
\def\expk{{\rm exp}\biggl[\,\sum_{i<j=1}^4 G_{Bij}k_i\cdot k_j\biggr]}
\def\expp{{\rm exp}\biggl[\,\sum_{i<j=1}^4 G_{Bij}p_i\cdot p_j\biggr]}
\def\expshort{{\e}^{\half G_{Bij}k_i\cdot k_j}}
\def\expabb{{\e}^{(\cdot )}}
\def\epseps#1#2{\varepsilon_{#1}\cdot \varepsilon_{#2}}
\def\epsk#1#2{\varepsilon_{#1}\cdot k_{#2}}
\def\kk#1#2{k_{#1}\cdot k_{#2}}
\def\G#1#2{G_{B#1#2}}
\def\Gp#1#2{{\dot G_{B#1#2}}}
\def\GF#1#2{G_{F#1#2}}
\def\Dab{{(x_a-x_b)}}
\def\Dsq{{({(x_a-x_b)}^2)}}
\def\PITD{{(4\pi T)}^{-{D\over 2}}}
\def\4piTD{{(4\pi T)}^{-{D\over 2}}}
\def\4piT4{{(4\pi T)}^{-2}}
\def\TintmD{{\dps\int_{0}^{\infty}}{dT\over T}\,e^{-m^2T}
    {(4\pi T)}^{-{D\over 2}}}
\def\Tintm4{{\dps\int_{0}^{\infty}}{dT\over T}\,e^{-m^2T}
    {(4\pi T)}^{-2}}
\def\Tintm{{\dps\int_{0}^{\infty}}{dT\over T}\,e^{-m^2T}}
\def\Tint{{\dps\int_{0}^{\infty}}{dT\over T}}
\def\np{n_{+}}
\def\nm{n_{-}}
\def\Np{N_{+}}
\def\Nm{N_{-}}
\newcommand{\slG}{{{\dot G}\!\!\!\! \raise.15ex\hbox {/}}}
\newcommand{\Gd}{{\dot G}}
\newcommand{\Gund}{{\underline{\dot G}}}
\newcommand{\Gdd}{{\ddot G}}
\def\GBd12{{\dot G}_{B12}}
\def\Dx{\dps\int{\cal D}x}
\def\Dy{\dps\int{\cal D}y}
\def\Dpsi{\dps\int{\cal D}\psi}
\def\dint#1{\int\!\!\!\!\!\int\limits_{\!\!#1}}
\def\ddtau{{d\over d\tau}}
\def\ie{\hbox{$\textstyle{\int_1}$}}
\def\iz{\hbox{$\textstyle{\int_2}$}}
\def\id{\hbox{$\textstyle{\int_3}$}}
\def\ldop{\hbox{$\lbrace\mskip -4.5mu\mid$}}
\def\rdop{\hbox{$\mid\mskip -4.3mu\rbrace$}}
%
\newcommand{\1}{{\'\i}}
\newcommand{\no}{\noindent}
\def\non{\nonumber}
\def\dps{\displaystyle}
\def\sy{\scriptscriptstyle}
\def\sy{\scriptscriptstyle}

\section{Introduction}
\label{sec:intro}

\begin{figure}[htbp]
\begin{center}
 \includegraphics[width=0.28\textwidth]{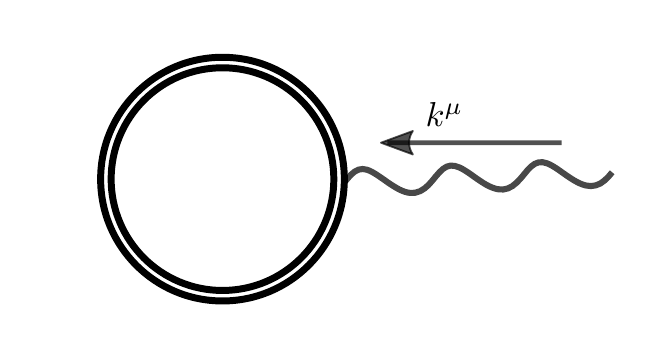}
\caption{{\bf One-loop one-photon amplitude in a constant external field.}}
\label{fig-tadpole}
\end{center}
\end{figure}

The QED one-loop one-photon amplitude vanishes in vacuum by Furry's theorem. 
In the presence of a constant external field this theorem does not imply that the one-photon
diagram (Fig. \ref{fig-tadpole}) vanishes. However, that diagram is still usually discarded, since it formally vanishes by momentum conservation
(as usual, the double-line denotes the electron propagator in a constant field). See, e.g., \cite{frgish-book}.

\begin{figure}[hb]
\begin{center}
 \includegraphics[width=0.3\textwidth]{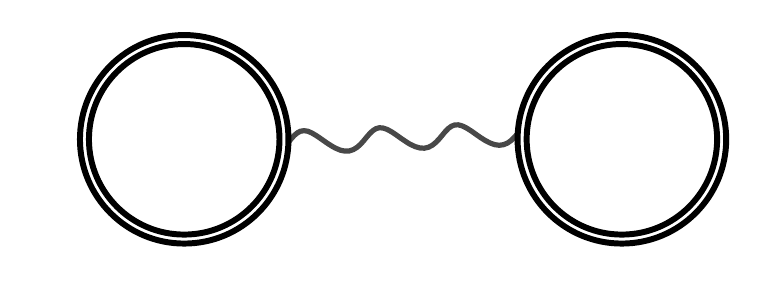}
\caption{{\bf One-particle reducible contribution to the two-loop EHL.}}
\label{fig-EHL1PR}
\end{center}
\end{figure}

\noindent Recently, Gies and Karbstein \cite{giekar} discovered that this diagram can cause non-vanishing contributions
when appearing as part of a larger diagram, due to the infrared singularity of the photon propagator connecting it 
to the rest of the diagram. As their principal example, they analyzed the one-particle reducible (`1PR') diagram in spinor QED 
shown in Fig. \ref{fig-EHL1PR}.

They showed that this diagram leads to a hitherto overlooked contribution ${\cal L}^{(2)\rm 1PR}$
to the two-loop Euler-Heisenberg Lagrangian \cite{eulhei,ritusspin,ditreuqed,18}.
Moreover, they found that this ``addendum'' can be written very
simply in terms of the one-loop Euler-Heisenberg Lagrangian ${\cal L}^{(1)}$:

\bear
{\cal L}^{(2)\rm 1PR}
&=& \partder{{\cal L}^{(1)}}{F^\mn} \partder{{\cal L}^{(1)}}{F_\mn} \, .
\label{gieskarb}
\ear
Two of the present authors in \cite{112} extended this result to scalar QED, and there also to the
open line case, i.e. to the scalar propagator in a constant field  $D_{\rm scal}(F)$. 
For the propagator, a 1PR diagram analogous to Fig. \ref{fig-EHL1PR} appears already at the one-loop level (Fig. \ref{fig-SE1PR}):

\begin{figure}[htbp]
\begin{center}
 \includegraphics[width=0.35\textwidth]{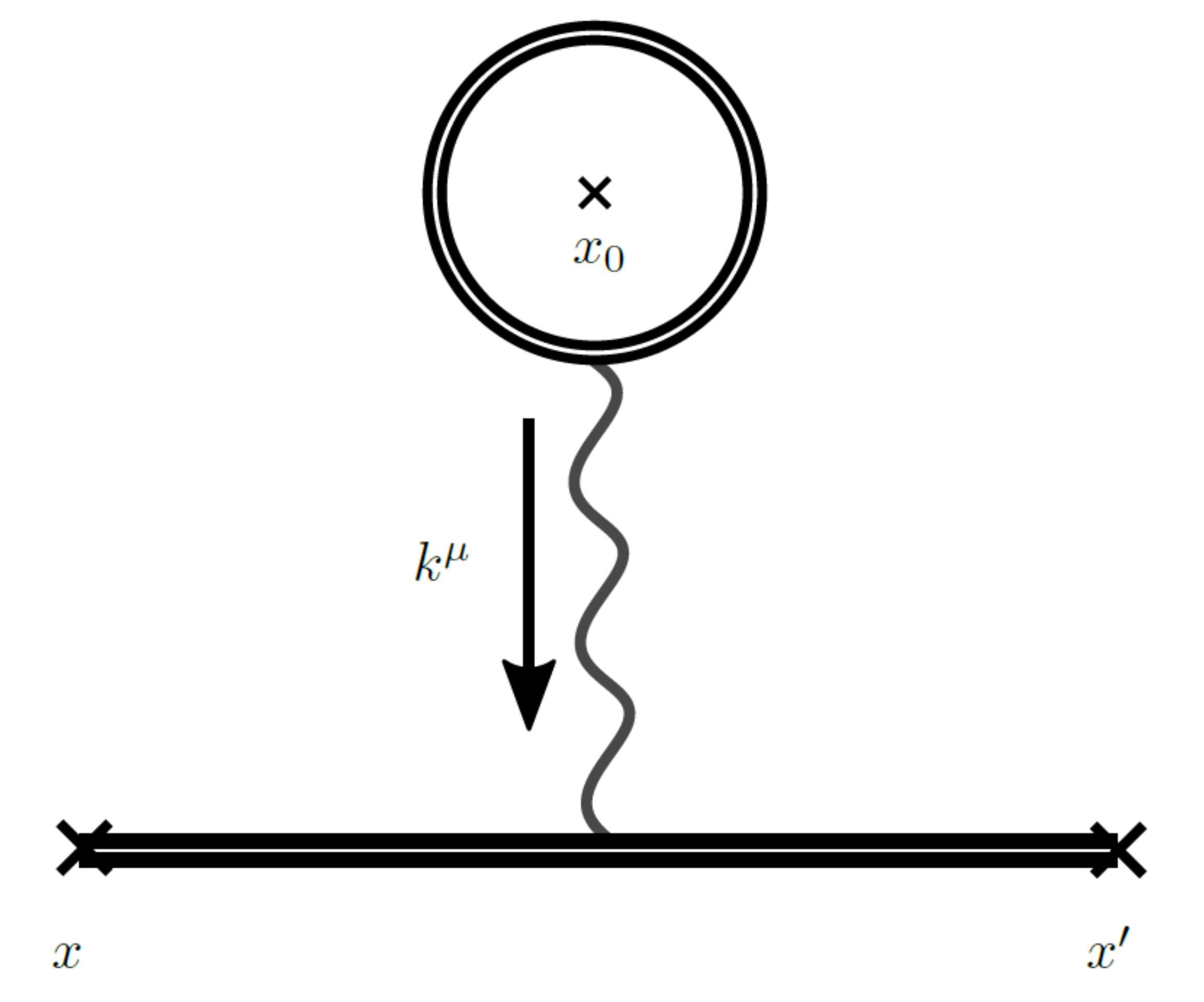}
\caption{{\bf 1PR contribution to the scalar propagator in a constant field.}}
\label{fig-SE1PR}
\end{center}
\end{figure}

\begin{figure}[htbp]
\begin{center}
 \includegraphics[width=0.35\textwidth]{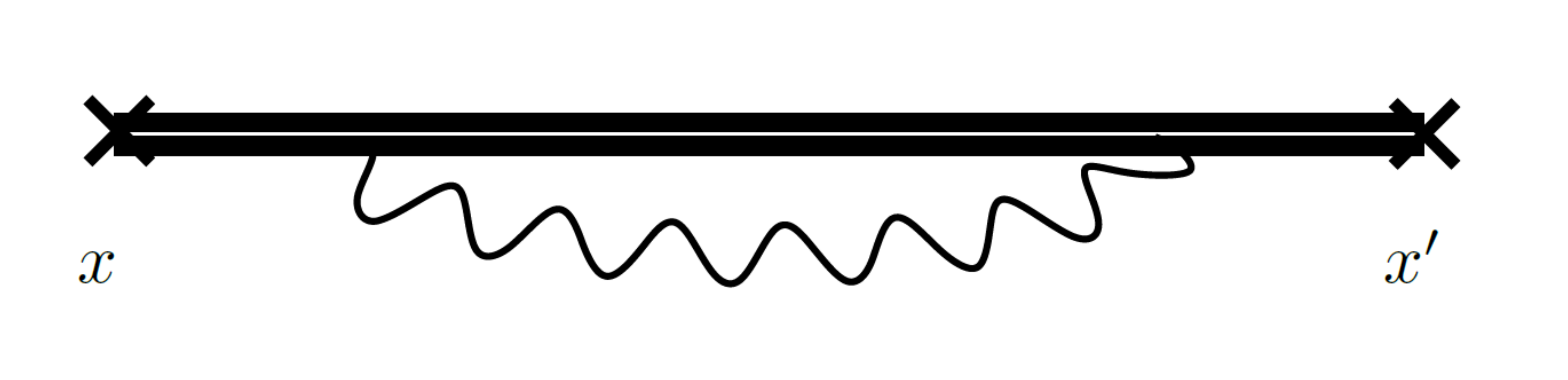}
\caption{{\bf One-loop propagator in a constant field.}}
\label{fig-SE1PI}
\end{center}
\end{figure}

\noindent
In \cite{112} it was found that equation \eqref{gieskarb} generalizes to the propagator in $x$-space as

\begin{equation}
D_{\rm scal}^{x'x(1)1PR} = 	\frac{\partial D^{x'x}_{\rm scal}}{\partial F_{\mu\nu}}\frac{\partial \mathcal{L}^{(1)}_{\rm scal}}{\partial F^{\mu\nu}} 
+ \frac{ie}{2}D^{x'x}_{\rm scal}x^{\prime\mu}\frac{\partial \mathcal{L}^{(1)}_{\rm scal}}{\partial F^{\mu\nu}}x^{ \nu}  \, .
	\label{resComp}
\end{equation}
Here Fock-Schwinger gauge centered at $x$ was chosen\footnote{We have interchanged $x$ and $x'$ with respect to the conventions of \cite{112}.}.
Of the two terms on the right-hand side, only the first one survives the Fourier transformation
to momentum space:

\begin{equation}
D_{\rm scal}^{(1)1PR}(p)  = 	\frac{\partial D_{\rm scal}(p)}{\partial F_{\mu\nu}}\frac{\partial \mathcal{L}^{(1)}_{\rm scal}}{\partial F^{\mu\nu}}  \, .
	\label{resCompp}
\end{equation}
This 1PR addendum is of the same order in $\alpha$ as the standard 1PI diagram shown in Fig. \ref{fig-SE1PI}.

The purpose of the present paper is to show that the equations \eqref{resComp}, \eqref{resCompp} generalize to the spinor QED case as they stand,
just replacing ${\cal L}_{\rm scal}, D_{\rm scal}$ by their spinor QED analogues, denoted by ${\cal L}$ and $S$ respectively. As in the scalar case, we will use the sewing relation 

\bear
S^{x'x(1)1PR}
&=& \int \frac{d^Dk}{(2\pi)^D k^2} 
\Gamma^{(1)}_{(1)}[k',\varepsilon';F]S^{x'x}_{(1)}[k,\varepsilon;F]
\Bigl\vert_{k'\to -k,\,\pol^{\mu}\pol'^{\nu}\to \eta^\mn}  
\nonumber\\
\label{sewingGammaD}
\ear
with the ingredients $\Gamma^{(1)}_{(1)}[k',\varepsilon';F]$ (the one-loop one-photon amplitude in the constant field)
and $S^{x'x}_{(1)}[k,\varepsilon;F]$ (the $x$-space spinor propagator in the field with one photon attached)
\footnote{By $S^{x'x}_{(N)}$ and $K^{x'x}_{(N)}$ we denote the constant field propagator and its kernel, the latter to be introduced below, with $N$ finite-energy photons attached 
(in addition to the zero-energy photons exchanged with the constant field background), and we refer to these objects as the ``$N$-photon propagator'' and ``$N$-photon kernel'', respectively.
Thus $N=0$ corresponds to the constant-field propagator itself, although we will keep writing $S^{x'x},K^{x'x}$ for it rather than $S^{x'x}_{(0)},K^{x'x}_{(0)}$.
In this context the constant-field propagator will also be called ``photonless'', despite of its interaction with the field.
}. 
The former contains a delta function $\delta^D(k)$ and one factor of momentum (see (\ref{idclosed}) below), so that by itself it vanishes. 
In the sewing relation \eqref{sewingGammaD} it will produce a finite contribution via

\bear
 \int d^Dk \, \delta^D(k) \frac{k^{\mu}k^{\nu}}{k^2} = \frac{\eta^\mn}{D} \, 
 \label{intk}
\ear
when combined with the part of $S^{x'x}_{(1)}[k,\varepsilon;F]$ that is also linear in the momentum (this may seem mathematically dubious,
but can be justified by regularizing the delta function as in \cite{giekar}, approaching the constant external field by a slightly space-time
dependent one, as is physically appropriate in any case). 

In \cite{112}, the identities  \eqref{resComp}, \eqref{resCompp} were shown by a direct calculation of all the ingredients,
using the worldline approach to QED in a constant field  \cite{strassler,5,15,shaisultanov,17,40,41,110}. Here, we will proceed
in a more efficient manner: rather than actually calculating the spinor propagator in the field and with one photon attached, which is the
only really new ingredient, we will write down a worldline path integral representation for this object, and show by manipulations under
the path integral that its linear part in the photon momentum $k^{\mu}$ -- which is all that is required for the sewing -- fulfills the identity
\begin{align}
S^{x'x}_{(1)}\Bigl\vert_k &=-2i\varepsilon \cdot \left(\frac{\partial S^{x'x}}{\partial F} + \frac{ie}{2}x_- S^{x'x} \xp \right) \cdot k -\frac{ie}{2} \varepsilon\cdot \big(\gamma \xm +\xm\gamma \big)\cdot k K^{x'x}
 \nonumber\\
&= -2i\varepsilon \cdot \left(\frac{\partial S^{x'x}}{\partial F} + \frac{ie}{2}x^{\prime} S^{x'x} x\right) \cdot k  +\varepsilon\cdot L \cdot k\,  ,
\label{idopen}	
\end{align}
where $x_+ \equiv \half (x+x')$, $x_- \equiv x' -x$, and
$L^{\mu\nu}$ is a symmetric tensor that will drop out upon sewing since the contribution from the 
one-loop one-photon diagram is anti-symmetric in its Lorentz indices (see \eqref{olop} below). Together with the similar identity for the closed loop \cite{112},

\bear
\Gamma^{(1)}_{(1)}[k,\varepsilon;F] 
= -2i (2\pi)^{D} \delta^D(k) 
\Bigl\lbrack\varepsilon\cdot\frac{\partial   {\cal L}^{(1)} (F)}{\partial F} \cdot k + O(k^3)\Bigr\rbrack\, , \nonumber\\
\label{idclosed}
\ear
it is then immediate to obtain, from the relations \eqref{sewingGammaD}, \eqref{intk},  the spinor QED generalization of  \eqref{resComp}, 

\begin{equation}
S^{x'x(1)1PR} = 	\frac{\partial S^{x'x}}{\partial F_{\mu\nu}}\frac{\partial \mathcal{L}^{(1)}}{\partial F^{\mu\nu}} 
+  \frac{ie}{2}S^{x'x}x^{\prime\mu}\frac{\partial \mathcal{L}^{(1)}}{\partial F^{\mu\nu}}x^{ \nu}  \, ,
	\label{resCompspin}
\end{equation}
and, by Fourier transformation, the momentum space version thereof,

\begin{equation}
S^{(1)1PR}(p)  = 	\frac{\partial S(p)}{\partial F_{\mu\nu}}\frac{\partial \mathcal{L}^{(1)}}{\partial F^{\mu\nu}}  \, .
	\label{resComppspin}
\end{equation}
The next section will be devoted to the demonstration of the ``derivative identity'' \eqref{idopen}. With the identities \eqref{resCompspin}, \eqref{resComppspin} in hand, 
we then use them in section \ref{sec:explicit} to write the 1PR contribution explicitly in $x$-space and momentum space, now requiring only the photonless versions of the
loop and propagator.  
In the concluding section we summarize our findings. 

\section{Worldline derivation of the derivative identity}

\subsection{The closed loop with zero and one photons}

As a warm-up, let us rederive the closed-loop formula \eqref{idclosed} in the new approach. 
In the worldline formalism, the one-loop spinor QED effective action can be written in terms of
a double worldline path integral as follows (see \cite{41} and refs. therein):

\begin{eqnarray}
\Gamma^{(1)}
\lbrack A\rbrack &  = &- \half {\displaystyle\int_0^{\infty}}
{dT\over T}
\e^{-m^2T}
{\displaystyle\int}_P 
D x
{\displaystyle\int}_A 
D\psi\nonumber\\
& \phantom{=}
&\times
{\rm exp}\biggl [- \int_0^T d\tau
\Bigl ({1\over 4}{\dot x}^2 + {1\over
2}\psi\cdot\dot\psi
+ ieA\cdot\dot x - ie\, \psi\cdot F\cdot\psi
\Bigr )\biggr ]
\, .
\label{spinorpi}
\end{eqnarray}
Here the orbital path integral $\int D x$ is over closed trajectories in space-time, $x(T) = x(0)$,
the spin path integral $\int D\psi$ over Grassmann functions $\psi^{\mu}(\tau)$
obeying antisymmetric boundary conditions, $\psi^{\mu}(T) = - \psi^{\mu}(0)$. 

For a constant $F_\mn$ it is convenient to use Fock-Schwinger gauge centered at the loop center of mass $x_0$ \cite{5}, 
since this allows one to write $A_{\mu}$ in terms of $F_{\mu\nu}$:

\bear
A^{\mu}(x) = \frac{1}{2} (x-x_0)^{\nu}F_{\nu\mu} \, .
\label{FS}
\ear
Separating off the loop center of mass via $x(\tau) = x_0 + q(\tau)$,
one obtains 

\bear
\Gamma^{(1)}(F) = \int d^D x_0 {\cal L}^{(1)} (F) \, ,
\label{GammatoL}
\ear
where

\begin{eqnarray}
{\cal L}^{(1)} (F) 
&  = &- \half {\displaystyle\int_0^{\infty}}
{dT\over T}
\e^{-m^2T}
{\displaystyle\int}_P 
D q
{\displaystyle\int}_A 
D\psi\nonumber\\
& \phantom{=}
&\times
{\rm exp}\biggl [- \int_0^T d\tau
\Bigl ({1\over 4}{\dot q}^2 + {1\over
2}\psi\cdot\dot\psi
+ \frac{ie}{2} q\cdot  F\cdot\dot q - ie \,\psi\cdot F\cdot\psi
\Bigr )\biggr ]
\label{spinorpiq}
\end{eqnarray}
and $q^{\mu}(\tau)$ now obeys the ``string-inspired'' constraint $\int_0^Td\tau q^{\mu}(\tau) =0$. 

The one-loop one-photon amplitude $\Gamma^{(1)}_{(1)}[k,\varepsilon;F]$  
is obtained from \eqref{spinorpi} by the insertion of the photon vertex operator

\bear
-ieV[k,\varepsilon]
=
-ie\int_0^Td\tau
\Bigl[
\varepsilon\cdot \dot x
+2i
\varepsilon\cdot\psi
k\cdot\psi
\Bigr]
\,\e^{ik\cdot x}
= -ie \e^{ik\cdot x_0} 
\int_0^Td\tau
\Bigl[
\varepsilon\cdot \dot q
+2i
\varepsilon\cdot\psi
k\cdot\psi
\Bigr]
\,\e^{ik\cdot q}
\nonumber\\
\label{photonvertop}
\ear
\noindent
under the path integrals in (\ref{spinorpiq}). Eventually the prefactor $\e^{ik\cdot x_0}$ integrating over $x_{0}$ will provide a momentum conserving $\delta(k)$,
which means that for the remaining $\tau$ - integral we are interested only in the leading, linear term in the momentum expansion (the integrand contains also a
momentum-independent term, which however integrates to zero).   
Projecting that part of the vertex operator correspondingly, we have

\bear
-ieV[k,\varepsilon]
\longrightarrow
\e^{ik\cdot x_0} 
\int_0^Td\tau \, e
\Bigl[
\varepsilon\cdot \dot q k\cdot q
+2 \varepsilon\cdot\psi k\cdot\psi
\Bigr]
\, .
\ear
We observe that, when acting on the exponent in \eqref{spinorpiq}, we can further replace the integrand in this last expression under the path integral by

\bear
-2i \e^{ik\cdot x_0}  \varepsilon^{\mu}\frac{\partial}{\partial F^{\mn}} k^{\nu} .
\ear
Putting things together we obtain eq. \eqref{idclosed},

\bear
\Gamma^{(1)}_{(1)}[k,\varepsilon;F] = -2i \int d^D x_0 \e^{ik\cdot x_0} \varepsilon^{\mu}\frac{\partial}{\partial F^{\mn}} k^{\nu}  {\cal L}^{(1)} (F)
= -2i (2\pi)^{D} \delta^D(k) \varepsilon\cdot\frac{\partial}{\partial F} \cdot k \, {\cal L}^{(1)} (F). \nonumber \\
\label{olop}
\ear

\subsection{The open line with zero and one photons}

The worldline representation of the spinor propagator in a constant field, dressed with photons, is a much more complicated
issue, and a formalism suitable for practical calculations has been developed only very recently \cite{114}, based on
\cite{18, fragit}. For the photonless case,
it leads to the representation\footnote{Our field theory conventions follow \cite{srednicki-book}, except that we use a
different sign of the elementary charge.}

\begin{align}
	S^{x'x}(F)  &= \left[m + i \gamma \cdot \left(\frac{\partial}{\partial x'} - \frac{ie}{2}F \cdot x_-\right)\right]
	K^{x'x}(F) 
	\label{D0photon}
\end{align}
with the ``kernel'' function

\begin{align}
	K^{x'x}(F)& =\int_0^\infty \e^{-m^2 T} dT\int_{x(0)=x}^{x(T)=x'}Dx\, \e^{-\int_0^Td\tau \big ({1\over 4}{\dot x}^2 
+ ieA\cdot\dot x \big)   }\nonumber\\
& \quad  \times \frac{1}{4} {\rm symb}^{-1}\Biggl\{\int_{A}D\psi\, \e^{-\int_0^Td\tau \big [ {1\over
2}\psi\cdot\dot\psi
 - ie\, \big(\psi+\frac12 \eta\big)\cdot F\cdot\big(\psi +\frac12 \eta\big)
\big]} \Biggr\} 
	\nonumber\\
&=\int_0^{\infty}
dT\,
\e^{-m^2T}
\e^{-\frac{x_-^2}{4T}}
\int
Dq\,
\e^{-\int_0^Td\tau\frac{1}{4}q\big(-\frac{d^2}{d\tau^2}+2ieF\frac{d}{d\tau}\big)q+\frac{ie}{T}x_-FQ}
\nonumber\\
& \quad \times \frac{1}{4}
{\rm symb}^{-1}
\int D\psi
\, \e^
{-\int_0^Td\tau\,
\bigl[\half\psi\big(\frac{d}{d\tau}-2ieF\big)\psi-ie\psi\cdot F\cdot \eta-\frac{ie}{4}\eta\cdot F\cdot \eta\bigl]
} \nonumber \\
	&= \int_{0}^{\infty} dT \e^{-m^{2}T} (4\pi T)^{-\frac{D}{2}} \detZs 
	\e^{-\frac{1}{4T} \xm \cdot \Zz \cdot \cot \Zz \cdot \xm}  \textrm{\rm symb}^{-1}\Bigl[\e^{\frac{i}{4} \eta \cdot \tan \Zz \cdot \eta}\Bigr] \, .
\label{K0photon}
\end{align}
Here the propagation is from $x$ to $x'$, $\eta^{\mu}$ is a constant Grassmann vector 
and ${\cal Z}_{\mu\nu} \equiv eF_\mn T$. 
Again we have used Fock-Schwinger gauge, now centered at the initial point $x$ (it is well-known how to convert the propagator from this gauge to a general gauge  \cite{ditgie-book}). In the second line we have introduced the orbital path-integral $\int Dq$ which runs over fluctuations about the straight-line path leading from
$x$ to $x'$; that is, trajectories $q(\tau)$ obeying Dirichlet boundary conditions in proper-time, $q(0)=q(T)=0$, and have defined 

\bear
Q^\mu\equiv \int_0^Td\tau \,q^\mu(\tau) \, .
\label{defQ}
\ear
The ``symbol map'' ${\rm symb}$  is defined by 

\bear
{\rm symb} 
\bigl(\hat\gamma^{[\alpha\beta\cdots\rho]}\bigr) \equiv 
\eta^\alpha\eta^\beta\ldots\eta^\rho
\label{defsymb}
\ear
where 

\bear
\hat\gamma^{\mu} \equiv i\sqrt{2} \gamma^{\mu}
\label{defhatgamma}
\ear
and $\hat\gamma^{[\alpha\beta\cdots\rho]}$ denotes the totally antisymmetrized product, 

\bear
\hat\gamma^{[\alpha_1\alpha_2\cdots \alpha_n]} \equiv \frac{1}{n!}\sum_{\pi\in S_n} {\rm sign}(\pi) \hat\gamma^{\alpha_{\pi(1)}}\hat\gamma^{\alpha_{\pi(2)}} \cdots \hat\gamma^{\alpha_{\pi(n)}} \, .
\label{Defantisymm}
\ear
By simple algebra, one may verify that

\bear
{\rm symb}^{-1}\Big[\e^{\frac{i}{4}\eta\cdot \tan\cZ\cdot\eta}\Big]
&=&\Eins-\frac{i}{4}(\tan\cZ)^{\mu\nu}[\gamma^\mu,\gamma^\nu]+\frac{i}{8}\epsilon^{\mu\nu\alpha\beta}(\tan\cZ)^{\mu\nu}(\tan\cZ)^{\alpha\beta}\gamma_5
\nonumber\\
\label{symbexp}	
\ear
and it is then easy to identify \eqref{D0photon} with the form given, e.g., in \cite{ditgie-book}.

Similarly, the fermion propagator with one photon attached can be written as \cite{114}

\bear
S^{x'x}_{(1)}[k,\varepsilon;F]&=& \left[m + i \gamma \cdot \left(\frac{\partial}{\partial x'} - \frac{ie}{2}F \cdot x_- \right)\right]  K_{(1)}^{x'x}[k,\varepsilon;F]
-e\slash\varepsilon \e^{ik\cdot x'}K^{x'x}(F)
\, .
\nonumber\\
\label{StoK}
\ear
Here the photonless kernel $K^{x'x}(F)$ has been given in \eqref{K0photon} above, and the one-photon kernel $K^{x'x}_{(1)}[k,\varepsilon;F]$
has the following path-integral representation:

\bear
\hspace{-1cm}K^{x'x}_{(1)}[k,\varepsilon;F] &=& -\frac{ie}{4}
\int_0^{\infty}
dT\,
\e^{-m^2T}
\e^{-\frac{x_-^2}{4T}}
\int
Dq\,
\e^{-\int_0^Td\tau\frac{1}{4}q\big(-\frac{d^2}{d\tau^2}+2ieF\frac{d}{d\tau}\big)q+\frac{ie}{T}x_-FQ}
\nonumber\\
&& \times 
{\rm symb}^{-1}
\int D\psi
\, \e^
{-\int_0^Td\tau\,
\bigl[\half\psi\big(\frac{d}{d\tau}-2ieF\big)\psi-ie\psi\cdot F\cdot \eta-\frac{ie}{4}\eta\cdot F\cdot \eta\bigl]
}V^{x'x}_{\eta}[k,\varepsilon]. \nonumber\\
\label{Kconst}
\ear
The photon vertex operator now appears in the form 

\bear
 V^{x'x}_{\eta}[k,\varepsilon] 
&=&   \int_0^Td\tau \biggl\lbrack \varepsilon\cdot \Bigl( \frac{x_-}{T} + \dot q\Bigr) 
+ 2i\varepsilon\cdot \Bigl(\psi+ \frac{\eta}{2}\Bigr) k\cdot \Bigl(\psi+ \frac{\eta}{2}\Bigr)
\biggr\rbrack \e^{ik\cdot \bigl( x+ x_-\frac{\tau}{T} + q\bigr)} 
\, .
 \nonumber\\
\label{vertexqpsi}
\ear
Again we are free to restrict this operator to its linear part,

\begin{equation}
	 V^{x'x}_{\eta}[k,\varepsilon]\link
= i\varepsilon \cdot \int_0^T d\tau \biggl\lbrack \Bigl( \frac{x_-}{T} + \dot q\Bigr) \left(x + \frac{x_-}{T}\tau + q\right)
+ 2\Bigl(\psi+ \frac{\eta}{2}\Bigr)  \Bigl(\psi+ \frac{\eta}{2}\Bigr)
\biggr\rbrack  \cdot k \, .
\label{vertexK}
\end{equation}
With suitable integration by parts and application of the boundary conditions this can be written as
\bear
	 V^{x'x}_{\eta}[k,\varepsilon]\link
= i\varepsilon \cdot  \biggl\lbrack  \frac{1}{2}x_-(x + x^{\prime}) + \int_0^T d\tau \Bigl( \dot{q} q + \frac{x_-}{T}q - q \frac{x_-}{T}	
+ 2\Bigl(\psi+ \frac{\eta}{2}\Bigr)  \Bigl(\psi+ \frac{\eta}{2}\Bigr) \Bigr)
\biggr\rbrack  \cdot k \, . \nonumber\\
\label{Vsimp}
\ear
In this form it is easy to see that an insertion of $-ieV^{x'x}_{\eta}[k,\varepsilon]\link$ into the $N=0$ kernel is equivalent to 
acting on it with the operator

\bear
(-2i)\varepsilon \cdot \left( \frac{\partial}{\partial F} + \frac{ie}{2}\xm \xp \right) \cdot k \, ,
\ear
leading to

\bear
K^{x'x}_{(1)}[k,\varepsilon;F]\Big|_k=(-2i)\varepsilon \cdot \left( \frac{\partial}{\partial F} + \frac{ie}{2}\xm \xp \right) \cdot k \, K^{x'x}(F)
\, .
\ear
Using this derivative identity for $K^{x'x}_{(1)}$ in \eqref{StoK}, and expanding the second term of that equation to linear order in $k$, it is then simple to show the derivative identity for $S^{x'x}_{(1)}$ itself, eq.~\eqref{idopen}. As described in the introduction this immediately leads to our compact expression for the addendum\footnote{We remark that $S^{x'x}_{(1)}[k,\varepsilon;F]$ also contains a term independent of $k$, but it does not contribute in the sewing procedure.
The reason is that it leads to a $k$-integral that vanishes by antisymmetry when matched with the lowest-order term in $\Gamma^{(1)}_{(1)}[k',\varepsilon';F]$, and,
as  indicated in \eqref{idclosed}, the next-higher term in the momentum expansion of  $\Gamma^{(1)}_{(1)}[k',\varepsilon';F]$ is already of order $\delta^D(k')k'^3$ \cite{112}.} in the spinor case, equation \eqref{resCompspin}. 

\section{Explicit form of the 1PR addendum}
\label{sec:explicit}

After having proven the identities \eqref{resCompspin}, \eqref{resComppspin}, we can now use them to forget about the
one-photon amplitudes, and work out the 1PR contribution from the photonless propagator, as given in \eqref{D0photon},
and the one-loop Euler-Heisenberg Lagrangian ${\cal L}^{(1)}(F)$. A representation of the latter suitable for our purposes is \cite{5}

 \bear
{\cal L}^{(1)}(F) &=&  
-2\int_0^{\infty}\frac {dT}{T}
(4\pi T)^{-{D\over 2}}
\e^{-m^2T}
{\rm det}^{-{1\over 2}}
\biggl[{{\rm tan}{\cal Z}\over {\cal Z}}\biggr].
\nonumber\\
\label{L1spinF}
\ear

\subsection{The 1PR addendum in configuration space}
In configuration space, an explicit representation of the addendum can be found by simply carrying out the differentiation (note that all matrices are built out of the constant anti-symmetric field strength tensor and so commute with one another) of the photonless propagator and one-loop effective action:

\bear
S^{x'x(1)1PR}&=&\e^{2}\int_{0}^{\infty} dT T\,\e^{-m^{2}T}(4\pi T)^{-\frac{D}{2}}\textrm{det}^{-\frac{1}{2}}\left[ \frac{\tan \Zz}{\Zz} \right]\e^{-\frac{1}{4T}\xm \Zz \cdot \cot \Zz \cdot \xm}
	\nonumber\\
&&\times
\int_{0}^{\infty}dT^{\prime}(4\pi T^{\prime})^{-\frac{D}{2}}\e^{-m^{2}T^{\prime}}\textrm{det}^{-\frac{1}{2}}\left[ \frac{\tan \Zzp}{\Zzp} \right]
 \bigg\lbrace  \Bigl\lbrack m - \frac{i}{2T}\gamma \cdot \Zz \cdot \left(  \cot \Zz  + i \right)\cdot \xm\Bigr\rbrack\nonumber\\
&&\times \bigg[ \frac{i}{2T}x^{\prime} \cdot \Xi^{\prime} \cdot x - \frac{1}{4T}\xm \cdot \left(\cot \Zz - \Zz \cdot \csc^{2}\Zz\right)\cdot \Xi^{\prime} \cdot \xm +   \frac{1}{2}\tr (\Xi \cdot \Xi^{\prime}) 
+ \Xi^{\prime}_{\mn}\frac{\partial}{\partial {\cal Z}_{\mn}}
\bigg]  \nonumber \\
&&- \frac{i}{2T} \gamma \cdot \left(\cot \Zz - \Zz \cdot \csc^{2}\Zz + i\right)\cdot \Xi^{\prime}\cdot \xm   \biggr\rbrace
 \textrm{symb}^{-1}\Big\{ \e^{\frac{i}{4} \eta \cdot \tan \Zz \cdot \eta} \Big\}
 \, .
\label{x-fin}
\ear
Here we have abbreviated

\bear
\Xi \equiv  \frac{1}{\sin \Zz \cdot \cos \Zz} - \frac{1}{\Zz}   = \frac{d}{d\Zz} \ln  \left[ \frac{\tan \Zz}{\Zz} \right],
\label{defXi}
\ear
and accordingly for $\Xi^{\prime}$. It remains to apply the inverse symbol map, which requires the identity \eqref{symbexp} and one further relation,

\begin{align}
\hspace{-2em}	
 \Xi^{\prime}_{\mn}\frac{\partial}{\partial {\cal Z}_{\mn}}
\textrm{symb}^{-1}\Bigl\lbrace \e^{\frac{i}{4}\eta \cdot \tan \Zz \cdot \eta} \Bigr\rbrace &= \textrm{symb}^{-1}\bigg\{ \frac{i}{4}\eta \cdot \sec^{2}\Zz \cdot \Xi^{\prime} \cdot \eta \,\e^{\frac{i}{4} \eta \cdot \tan \Zz \cdot \eta} \bigg\}\nonumber \\
	& = -\frac{i}{4}\bigg[\left(\sec^{2}\Zz \cdot \Xi^{\prime}\right)_{\mu\nu} [\gamma^{\mu}, \gamma^{\nu}] 
	- \epsilon^{\mu\nu\alpha\beta} \left(\sec^{2}\Zz \cdot\Xi^{\prime}\right)_{\mu\nu} (\tan\Zz)_{\alpha\beta} \gamma_{5}\bigg].
	\nonumber\\
	\label{symbdF}
\end{align} 

\subsection{The 1PR addendum in momentum space}

The momentum space version of the addendum can be obtained either by Fourier transforming our $x$ -space result above \eqref{x-fin}, or by a direct use of the momentum space derivative
identity \eqref{resComppspin}. To give an explicit expression, first we need the photonless propagator in momentum space \cite{114}

\bear
	S(p) &=& \int_{0}^{\infty} dT \, \bigl[m - \gamma \cdot (\Eins + i{\tan \Zz}) \cdot p\bigr] \e^{-T(m^{2} + p \cdot \frac{\tan \Zz}{\Zz} \cdot p)} \textrm{symb}^{-1}\bigg \{ \e^{\frac{i}{4} \eta \cdot \tan \Zz \cdot \eta} \bigg\}
	\, .
	\nonumber\\
\ear
After simple algebra, one finds

\bear
S^{(1)1PR}(p)&=&\e^{2}\int_{0}^{\infty} dT T\,\e^{-T(m^{2} + p \cdot \frac{\tan \Zz}{\Zz} \cdot p)}\int_{0}^{\infty}dT^{\prime}(4\pi T^{\prime})^{-\frac{D}{2}}\e^{-m^{2}T^{\prime}}
\textrm{det}^{-\frac{1}{2}}\Bigl\lbrack\frac{\tan \Zzp}{\Zzp} \Bigr\rbrack\nonumber \\
&&\times \bigg\lbrace\Bigl\lbrack m - \gamma \cdot(\Eins + i{\tan{\Zz}}) \cdot p  \Bigr\rbrack
\bigg[	- T p \cdot \frac{\Zz - \sin \Zz \cdot \cos \Zz}{\Zz^{2} \cdot \cos^{2}\Zz} \cdot \Xi^{\prime} \cdot p +  \Xi^{\prime}_{\mn}\frac{\partial}{\partial {\cal Z}_{\mn}} \bigg] 
\nonumber\\
&&\,\,\, - i \gamma \cdot \sec^{2}\Zz \cdot
\Xi^{\prime} \cdot p 
\bigg\rbrace
\textrm{symb}^{-1} \bigg\{ \e^{\frac{i}{4}\eta \cdot \tan \Zz \cdot \eta}\bigg\} 
\, ,
\ear
where the same relations (\ref{symbexp}) and (\ref{symbdF}) are required to apply the inverse symbol map. 
This concludes the determination of the addendum for the spinor propagator.

\section{Summary and outlook}
Extending recent results we have shown that there is a one-particle reducible contribution to the spinor propagator in a constant field at one-loop order,
given by a Feynman diagram that was previously believed to vanish. This contribution is of the same order in $\alpha$ as the standard 1PI self-energy
diagram in QED in a constant field. We have derived formulas for this addendum in terms of the spinor propagator in the field, and the one-loop
Euler-Heisenberg Lagrangian. Although these formulas are analogous to the ones obtained for scalar QED in \cite{112}, here we have introduced a
more efficient approach to their derivation in the framework of the worldline formalism, using formal manipulations under the path integrals rather
than an actual calculation of the path integrals as had been done in \cite{112}. 
We have given explicit but compact expressions for the addendum in configuration space and momentum space. 
These expressions should make it easy to study the effect of the addendum on such important quantities in external-field QED as the 
strong-field asymptotics \cite{kuzmik-book} and the Ritus mass shift \cite{ritusmass}. 

As in the case of the Euler-Heisenberg Lagrangian studied in \cite{giekar}, here, too, for possible applications of the addendum
it must be stressed that the 1PR diagram, unlike the 1PI one, can 
 involve two different fermions, e.g. we could have an electron in the loop and a quark in the line, or vice versa. 

At higher loop order the non-vanishing of these lately uncovered contributions will clearly lead to a proliferation of previously overlooked terms of this type
in constant-field QED.

\subsection*{Acknowledgements}

The work of N.A. was supported by IBS (Institute for Basic Science) under grant No. IBS-R012-D1. J.P.E. and C.S. thank CONACYT for financial support through 
grant Ciencias Basicas 2014 No. 242461.


\end{document}